\documentclass[twocolumn,showpacs,preprintnumbers,amssymb]{revtex4}

\usepackage{graphicx}
\usepackage{bm}
\begin{document}

\centerline{}
\title{Highly Damped Quasinormal Modes of Kerr Black Holes:\\
A Complete Numerical Investigation}
\author{Emanuele Berti}
\email{berti@iap.fr}
\affiliation{
McDonnell Center for the Space Sciences, Department of Physics,
Washington University, St. Louis, Missouri 63130, USA.
\footnote{Present address: Groupe de Cosmologie et Gravitation
(GR$\epsilon$CO), Institut d'Astrophysique de Paris (CNRS), $98^{bis}$
Boulevard Arago, 75014 Paris, France}}
\author{Vitor Cardoso}
\email{vcardoso@fisica.ist.utl.pt}
\affiliation{
Center for Computational Physics, University of Coimbra, P-3004-516
Coimbra, Portugal
\footnote{Present address: Centro Multidisciplinar de Astrof\'{\i}sica
- CENTRA, Departamento de F\'{\i}sica, Instituto Superior T\'ecnico,
Av. Rovisco Pais 1, 1049-001 Lisboa, Portugal}}
\author{Shijun Yoshida}
\email{yoshida@fisica.ist.utl.pt}
\affiliation{
Centro Multidisciplinar de Astrof\'{\i}sica - CENTRA, 
Departamento de F\'{\i}sica, Instituto Superior T\'ecnico,
Av. Rovisco Pais 1, 1049-001 Lisboa, Portugal}

\date{\today}

\begin{abstract}
We compute for the first time very highly damped quasinormal modes of
the (rotating) Kerr black hole. Our numerical technique is based on a
decoupling of the radial and angular equations, performed using a
large-frequency expansion for the angular separation constant
$_{s}A_{l m}$. This allows us to go much further in overtone number
than ever before. We find that the real part of the quasinormal
frequencies approaches a non-zero constant value which does {\it not}
depend on the spin $s$ of the perturbing field and on the angular
index $l$: $\omega_R=m\varpi(a)$. We numerically compute
$\varpi(a)$. Leading-order corrections to the asymptotic frequency are
likely to be $\sim 1/\omega_I$. The imaginary part grows without
bound, the spacing between consecutive modes being a monotonic
function of $a$.
\end{abstract}


\maketitle
\newpage
\section{Introduction}
Black holes (BHs), as many other objects, have characteristic
vibration modes, called quasinormal modes (QNMs). The associated
complex quasinormal frequencies (QN frequencies) depend only on the BH
fundamental parameters: mass, charge and angular momentum. QNMs are
excited by BH perturbations (as induced, for example, by infalling
matter). The early evolution of a generic perturbation can be
described as a superposition of QNMs, and the characteristics of
gravitational radiation emitted by BHs are intimately connected to
their QNM spectrum.  One may in fact infer the BH parameters by
observing the gravitational wave signal impinging on the detectors
\cite{echeverria}: this makes QNMs highly relevant in the newly born
gravitational wave astronomy \cite{schutz,kokkotas}.

Besides this ``classical'' context, QNMs may find a very important
place in the realm of a quantum theory of gravity. General
semi-classical arguments suggest \cite{bekenstein} that on quantizing
the BH area one gets an evenly spaced spectrum of the form
\begin{equation}
A_n=4\log{(k)}\,l_{P}^2 n\,\,;\,\, n=0,1,...
\label{areaspectrum}
\end{equation}
where $l_{P}$ is the Planck length, and $k$ is an integer to be
determined.  Hod \cite{hod} proposed to fix the value of $k$, and
therefore the area spectrum, by promoting QN frequencies with a very
large imaginary part to a special position: they should bridge the gap
between classical and quantum transitions.  Hod obtained, for the
Schwarzschild BH, $k=3$.  Following his proposal, further enhanced by
the prospect of using similar reasoning in Loop Quantum Gravity to fix
the Barbero-Immirzi parameter \cite{dreyer}, the interest in highly
damped BH QNMs has grown considerably \cite{cardosolemosshijun}. There
is now reason to believe that the connection between QN frequencies
and the BH area quantum is not as straightforward as initially
suggested. However a relation between classical and quantum BH
properties does seem to exist, even in non-asymptotically flat
spacetimes \cite{carlip}. A prerequisite to study this connection is
to compute QN frequencies having very large imaginary part. So far
this problem has been solved only for a few special geometries:
Schwarzschild BHs \cite{nollert,motl1,motl2,andersson,bertikerr1},
Reissner-Nordstr\"om (RN) BHs \cite{motl2,bertikerr1,andersson}, the
Ba\~nados-Teitelboim-Zanelli BH \cite{cardosobtz}, and the
four-dimensional Schwarzschild-anti-de Sitter BH \cite{cardosoroman}.

We must try to include the Kerr geometry in this short catalogue, due
to its great importance and simplicity.  This is a problem of great
relevance for the scientific community, and quite a lot of effort is
being invested here. This effort is in direct proportion to the
difficulty of the problem.  All previous attempts
\cite{kerranalytical,hisashi,bertikerr1,bertikerr2} at probing the
asymptotic QNMs of Kerr BHs have been unsuccessful, or at least
unsatisfactory.  There have been several contradictory ``analytical''
results, which were either based on incorrect assumptions, or could
not probe the highly damped regime \cite{kerranalytical}.  The few
numerical results \cite{hisashi,bertikerr1,bertikerr2} are not
decisive either, although they definitely show some trend.  A
numerical investigation is necessary both as a benchmark and as a
guide to analytical approaches.  Here we carry out such a numerical
study. We improve on previous results by going further in overtone
number than ever before, in order to really probe the asymptotic
regime. The starting point for our analysis is, as previously
\cite{bertikerr1,bertikerr2}, Leaver's continued fraction technique
\cite{leaver} as improved by Nollert \cite{nollert}, with a few
appropriate modifications \cite{bertikerr1}. However, we now decouple
the angular and radial equations. We first determine the asymptotic
expansion for the angular separation constant, and then replace this
asymptotic expansion in the radial equations. This trick spares us the
need to solve simultaneously the two equations, which was the major
drawback of previous numerical works. A leading-order asymptotic
expansion of the separation constant is typically accurate for
$|a\omega|\gtrsim 5$, where $a$ is the dimensionless Kerr rotation
parameter \cite{conventions}. In this study we go well beyond this
regime (we can usually compute modes up to $|a\omega|>50$, an order of
magnitude larger).  So we have great confidence that our results
really yield ``asymptotic'' QN frequencies.

We find that our previous results \cite{bertikerr2} for negative $m$
and moderately damped QN frequencies were quite close to the true
asymptotic behaviour (especially for large values of $a$), while
convergence to the asymptotic value was not yet achieved for positive
$m$. Our improved calculations have been carried out with two
independent numerical codes. Our main results are that:
(i) The real part of the QN frequencies $\omega_R$ approaches a
non-zero constant value. This value does {\it not} depend on the spin
$s$ of the perturbing field and on the angular index $l$. It only
depends on the rotation parameter $a$, and is proportional to $m$:
$\omega_R=m\varpi(a)$. We determine $\varpi(a)$ numerically.  A fit of
our numerical data by power series in $1/\omega_I$ and
$\sqrt{1/\omega_I}$ suggests that leading-order corrections to the
asymptotic frequency should be of order $1/\omega_I$.
(ii) The imaginary part $\omega_I$ grows without bound, the spacing
between modes $\delta \omega_I$ being a monotonically increasing
function of $a$.

\section{Basic Equations}
In the Kerr geometry, the condition that a given frequency be a QN
frequency can be converted into a statement about continued fractions,
which is rather easy to implement numerically.  Such a procedure has
been explained thoroughly by Leaver \cite{leaver}, so here we shall
only recall the basic idea.  The perturbation problem reduces to a
pair of coupled differential equations: one for the angular part of
the perturbations, and the other for the radial part. In
Boyer-Lindquist coordinates, defining $u=\cos\theta$, the angular
equation reads
\begin{eqnarray}
\label{angularwaveeq}
&&\left [
(1-u^2)S_{lm,u}
\right]_{,u}+\\
\nonumber
&+&\left[
(a\omega u)^2-2a\omega su+s+_{s}A_{lm}-{(m+su)^2\over 1-u^2}
\right]
S_{lm}=0,
\end{eqnarray}
and the radial one is
\begin{equation}
\Delta R_{lm,rr}+(s+1)(2r-1)R_{lm,r}+V(r)R_{lm}=0,
\label{pot}
\end{equation}
where $\Delta=r^2-r+a^2$ and
\begin{eqnarray}
&&V(r)={\biggl [}
(r^2+a^2)^2\omega^2
+is(am(2r-1)-\omega(r^2-a^2))+
\nonumber
\\ 
&&+a^2m^2-2am\omega r
{\biggr ]}
\Delta ^{-1}
+
\left[
2is\omega r-a^2\omega^2-_{s}A_{lm}
\right].
\label{radial}
\end{eqnarray}
The parameter $s=0,-1,-2$ for scalar, electromagnetic and
gravitational perturbations respectively, and $_{s}A_{l m}$ is an
angular separation constant. In the Schwarzschild limit the angular
separation constant can be determined analytically:
$_{s}A_{lm}=l(l+1)-s(s+1)$.

Boundary conditions for the two equations can be cast as a couple of
three-term continued fraction relations \cite{leaver}.  Finding QN
frequencies is a two-step procedure: for assigned values of $s,l,m,a$
and $\omega$, first find the angular separation constant
$_{s}A_{lm}(\omega)$ looking for zeros of the {\it angular} continued
fraction; then replace the corresponding eigenvalue into the {\it
radial} continued fraction, and look for its zeros as a function of
$\omega$. This has been the strategy adopted in earlier works
\cite{bertikerr1,bertikerr2,hisashi,leaver}, where the first $\sim 50$
modes were computed. These numerical investigations showed a rich (and
perhaps confusing) behavior. For negative $m$ and large enough $a$ the
first $\sim 50$ modes display some kind of convergence, and are
consistent with our new results. Among modes with positive $m$, only
those having $|s|=l=m=2$ seemed to converge. This convergence was
deceiving: positive-$m$ results in \cite{bertikerr2} were not yet in
the asymptotic regime. The ``true'' asymptotic behavior turns out to
be much simpler than the intermediate-damping regime explored in
\cite{bertikerr2}.

Since the major numerical difficulty lies in the coupling of the two
continued fractions, here we adopt a ``trick'' to decouple them. We
first carry out a careful study of the angular equation, determining
the asymptotic form of the separation constant $_{s}A_{lm}$ for
frequencies $\omega$ with large imaginary part. Then we substitute
this expansion in Eqs. (\ref{pot})-(\ref{radial}). By this trick we
reduce the problem to the numerical solution of a {\it single}
three-term recursion relation: it is then possible to probe the
asymptotic regime for highly damped modes. In the following section we
shall briefly discuss our main analytical and numerical results for
the asymptotic expansion of $_{s}A_{lm}$.

\begin{figure}
\centerline{\includegraphics[width=6.5cm,angle=270]{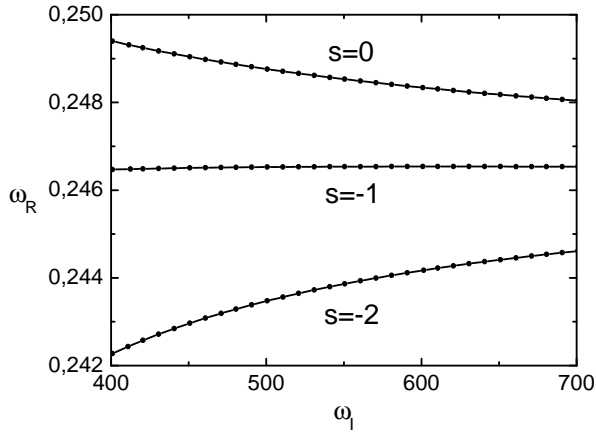}}
\caption{Real part of high-order QN frequencies for scalar ($s=0$),
electromagnetic ($s=-1$) and gravitational ($s=-2$) perturbations of a
Kerr BH with $a=0.1$ ($l=m=2$) as a function of their imaginary
part. QN frequencies of different spins converge to the same
value. For any kind of perturbation we are already deep in the region
of validity of the asymptotic expansion (\ref{alm}), since $|a\omega|
\sim 60$.}
\label{fig:Kerralls}
\end{figure}

\section{Asymptotic expansion of the angular separation constant}
The analytical properties of the angular equation
(\ref{angularwaveeq}) and of its eigenvalues have been studied by many
authors \cite{flammer,early,seidel,breuerbook,breuer}. Series
expansions of $_{s}A_{lm}$ for $|a \omega|\ll 1$ are available, and
they agree well with numerical results \cite{seidel}. On the other
hand, the asymptotic behavior for large frequencies has hardly been
studied at all. An analytical power-series expansion for large (pure
real and pure imaginary) values of $a \omega$ can be found in
Flammer's book \cite{flammer}, but it is limited to the case $s=0$.
Flammer's results are in good agreement with the exhaustive numerical
work by Oguchi \cite{oguchi}, who computed angular eigenvalues for
complex values of $a \omega$ and $s=0$. A review of numerical methods
to compute eigenvalues and eigenfunctions for $s=0$ can be found in
\cite{li}. Quite surprisingly, there are no systematic numerical
results for general spin $s$, and the few analytical predictions for
large values of $|a \omega|$ do not agree with each other
\cite{comment}. Here we shall fill this gap, presenting some results
for the large-$|a \omega|$ expansion of $_{s}A_{lm}$.

\begin{figure}
\centerline{\includegraphics[width=6.5cm,angle=270]{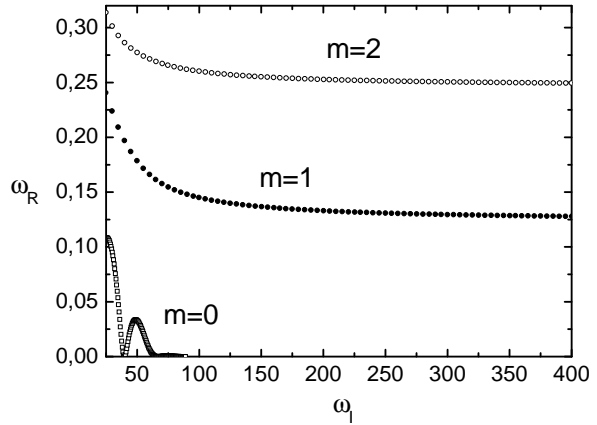}}
\caption{Scalar QN frequencies of a Kerr BH with $a=0.1$, $l=2$ and
different values of $m$. The asymptotic value is proportional to
$m$. Modes with $m=0$ oscillate around zero. The amplitude of these
oscillations at fixed $a$ decreases very fast, probably tending to
zero as $\omega_I\to \infty$: this is consistent with the behavior
shown in Fig. 6 of \cite{bertikerr2}.}
\label{fig:Kerrallm}
\end{figure}

A straightforward generalization of Flammer's method can be easily
found for general $s$. Define a new angular wavefunction $Z_{lm}(u)$
through \cite{breuerbook}
\begin{equation} 
S_{lm}(u)=(1-u^2)^{\frac{m+s}{2}}Z_{lm}(u)\,,
\label{newwave}
\end{equation} 
and change independent variable by defining $x=\sqrt{2c}{u}$, where
$c^2=-(a\omega)^2$.  Substitute this in (\ref{angularwaveeq}) to get:
\begin{eqnarray} 
&& \!\!\! \left[ _{s}A_{lm}-\frac{cx^2}{2}-i\sqrt{2c}x-m(m+1)-\frac{2msx}{\sqrt{2c}+x}\right] Z_{lm}+
\nonumber
\\
&+&(2c-x^2)Z_{lm,xx}-2(m+s+1)x Z_{lm,x}=0\,.
\label{angwaveeq2}
\end{eqnarray}
When $c \rightarrow \infty$, this equation becomes a parabolic cylinder function.
The arguments presented in \cite{flammer,breuerbook,breuer} lead to
\begin{equation}
_{s}A_{lm}=(2L+1)c+ {\cal O}(c^0) \,\,,\, c \rightarrow \infty \,,
\label{alm}
\end{equation}
where $L$ is the number of zeros of the angular wavefunction inside the domain.
One can show that
\begin{equation}
L=\left\{ \begin{array}{ll}
            l-|m|\,,   & {|m| \geq |s|},\\ 
            l-|s|\,,   &{|m|<|s|}. 
\end{array}\right.
\label{jdef}
\end{equation}
Higher order corrections in the asymptotic expansion can be obtained
as indicated in \cite{flammer}. However, we will not need them
here. We have verified Eq. (\ref{alm}) numerically, solving Leaver's
angular continued fraction for $_{s}A_{lm}$ as a function of the
complex parameter $a\omega$. Our numerical results (which will be
presented in detail elsewhere) are in excellent agreement with
previous work \cite{flammer,oguchi,li} for $s=0$. For any $s$ they
are consistent with equation (\ref{alm}) when $\omega_R\ll \omega_I$
and $|a\omega|$ is large.

\section{Numerical results}

To compute the asymptotic QN frequencies of the Kerr black hole we use
a technique similar to that described in \cite{nollert}. We fix a
value of the rotation parameter $a$. We first compute QN frequencies
for which $|a \omega|\sim 1$, so that formula (\ref{alm}) is only
marginally valid. This procedure is consistent with our previous
intermediate-damping calculations: for example, when we include terms
up to order $|a\omega|^{-2}$ in the asymptotic expansion for
$_{0}A_{lm}$ provided in \cite{flammer}, our new results for $a\simeq
0.1$ and $l=m=2$ match the results for the scalar case presented in
\cite{bertikerr2} at overtone numbers $20\lesssim N\lesssim 30$.  Then
we increase the overtone index $N$ (progressively increasing the
inversion index of our ``decoupled'' continued fraction). Finally, we
fit our numerical results by functional relations of the form:
\begin{equation}
\omega_R(N)=\omega_R+\omega_R^{(1)}\alpha+\omega_R^{(2)}\alpha^2+\omega_R^{(3)}\alpha^3,
\nonumber
\end{equation}
where $\alpha=1/\omega_I$ or $\alpha=\sqrt{1/\omega_I}$.  At variance
with the non-rotating case \cite{nollert}, fits in powers of
$1/\omega_I$ perform better, especially for small and large
$a$. However, both fits break down as $a\to 0$: the values of
higher-order fitting coefficients increase in this limit, so that
subdominant terms become as important as the leading order, and the
extraction of the asymptotic frequency $\omega_R$ becomes
problematic. The numerical behavior of subdominant coefficients
supports the expectation (which has not yet been verified
analytically) that subdominant corrections are
$a$-dependent. Therefore one has to be careful to the order in which
the limits $N\to \infty$, $a\to 0$ are taken
\cite{motl2,bertikerr1,bertikerr2}. These observations are consistent
with the fact that, in the RN case, the zero-charge limit of the
asymptotic QN frequency spectrum does not yield the asymptotic
Schwarzschild QN spectrum \cite{motl2,andersson,bertikerr1}.

\begin{figure}
\centerline{\includegraphics[width=6.5cm,angle=270]{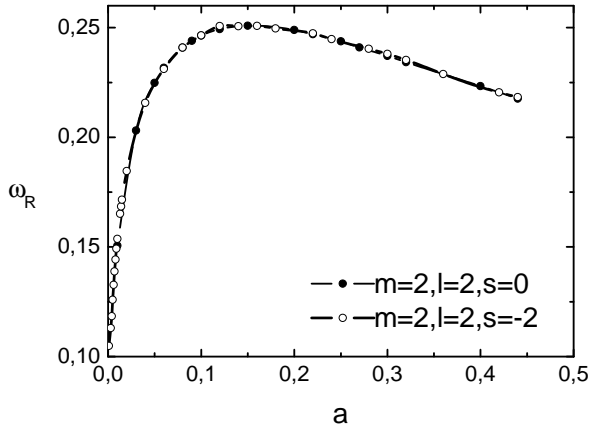}}
\caption{Asymptotic real part $\omega_R=2\varpi(a)$ of the $l=m=2$
gravitational and scalar QN frequencies extrapolated from numerical
data: $\omega_R\to 2\varpi(1/2)\simeq 0.21$ as $a\to 1/2$. The
extrapolated frequencies are independent on $l$ and $s$.}
\label{fig:Kerralla}
\end{figure}

We have extracted asymptotic frequencies using two independent
numerical codes.  For each value of $a$, we found that the
extrapolated value of $\omega_R$ is independent of $s$
(Fig. \ref{fig:Kerralls}), independent of $l$ and proportional to $m$
(Fig. \ref{fig:Kerrallm}): $\omega_R=m \varpi(a)$.

We obtained $\omega_R$ computing QN frequencies both for scalar
perturbations ($s=0$) and for gravitational perturbations
($s=-2$). For definiteness, in both cases we picked $l=m=2$. The
agreement between the extrapolated behaviors of $\omega_R$ as a
function of $a$ is excellent, suggesting that both sets of results are
typically reliable with an error $\lesssim 1 \%$.  Our results are
also weakly dependent on the number of terms used in the asymptotic
expansion of $_{s}A_{l m}$: this provides another powerful consistency
check. We have tried to fit the resulting ``universal function'',
displayed in Fig. \ref{fig:Kerralla}, by simple polynomials in the
BH's Hawking temperature $T$ and angular velocity $\Omega$ (and their
inverses). None of these fits reproduces our numbers with satisfactory
accuracy.  It is quite likely that asymptotic QN frequencies will be
given by an implicit formula involving the exponential of the Kerr
black hole temperature, as in the RN case
\cite{motl2,andersson,bertikerr1}.

For any $a$, the imaginary part $\omega_I$ grows without bound. Quite
surprisingly, the spacing between modes $\delta \omega_I$ is a
monotonically increasing function of $a$: it is not simply given by
$2\pi T$, as recent calculations and previous numerical results
suggested \cite{bertikerr1,spacing}. A power fit in $a$ of our
numerical results yields:
\begin{equation}
\delta \omega_I=1/2+0.0438a-0.0356a^2. 
\end{equation}

\section{conclusions}

Motivated by recent suggestions of a link between classical black hole
oscillations and quantum gravity, we have computed for the first time
very highly damped QNMs of the Kerr black hole. Our calculation was
made possible by a decoupling of the radial and angular equations,
carried out using asymptotic expansions of the angular separation
constant $_{s}A_{lm}$ for $\omega_r\ll \omega_I$ and $|a \omega|\gg
1$. Our results are very weakly dependent on the number of terms used
in the asymptotic expansion of $_{s}A_{l m}$, and this provides a
powerful consistency check. We found that:

(i) The real part of the QN frequencies $\omega_R$ approaches a
non-zero constant value.  This value does {\it not} depend on the spin
$s$ of the perturbing field and on the angular index $l$. It only
depends on the rotation parameter $a$, and is proportional to $m$:
\begin{equation}
\omega_R=m \varpi(a). 
\end{equation}
We determined $\varpi(a)$ numerically (Fig. \ref{fig:Kerralla}), and
showed that it is not given by simple polynomial functions of the
black hole temperature $T$ and angular velocity $\Omega$ (or their
inverses). At fixed $a$, a fit of our numerical data by power series
in $1/\omega_I$ and $\sqrt{1/\omega_I}$ suggests that leading-order
corrections to the asymptotic frequency are probably of order
$1/\omega_I$.
(ii) The imaginary part $\omega_I$ grows without bound, the spacing
between modes $\delta \omega_I$ being a monotonically increasing
function of $a$.

We wish to stress, once again, that the asymptotic frequency
$\omega_R$ is independent on the spin $s$ of the perturbing field:
this is consistent with results for highly damped QNMs of (charged) RN
black holes \cite{motl2,andersson}.

By now it is quite clear that the original Hod proposal requires some
modification. However, the ``universality'' of the asymptotic Kerr
behaviour we established in this paper is good news. For both charged
and rotating black holes the asymptotic QNM frequency $\omega_R$
depends only on the black hole geometry, not on the perturbing field.
If QNMs do indeed play a role in black hole quantization this is an
essential prerequisite, and it seems to hold.

\vskip 2mm

\section*{Acknowledgements}
We thank all members of the GR$\epsilon$CO group, Ted Jacobson, Kostas
Kokkotas, Lubos Motl, Andy Neitzke and Cliff Will for their interest
in this problem and many useful discussions. This work was partially
funded by Funda\c c\~ao para a Ci\^encia e Tecnologia (FCT) --
Portugal through project PESO/PRO/2000/4014.  VC acknowledges
financial support from FCT through PRAXIS XXI programme.  SY
acknowledges financial support from FCT through project SAPIENS
36280/99.



\begin{thebibliography}{99}
\bibitem{conventions} In this paper we adopt Leaver's conventions
\cite{leaver}.  In particular, we choose units such that $2M=1$.
\bibitem{comment} In \cite{breuerbook} an attempt was made to
generalize Flammer's large-$(a\omega)$ expansion for general spin
$s$. However, the result given there for the number of zeroes of the
angular wavefunction (page 115) is wrong (see Theorem 4.1 in
\cite{breuer}).  Therefore, although the derivation is essentially
correct, the expression for the leading term is not.
\end{thebibliography}
\end{document}